\documentclass[aps,prb,twocolumn,superscriptaddress,showpacs]{revtex4}
\usepackage{bm}
\usepackage{epsfig}
\usepackage{amsmath}
\usepackage{array}
\usepackage{lscape}
\renewcommand{\v}[1]{{\bf #1}}
\newcommand{\bpm}{\begin{pmatrix}}
\newcommand{\epm}{\end{pmatrix}}
\newcommand{\ba}{\begin{eqnarray}}
\newcommand{\ea}{\end{eqnarray}}
\newcommand{\nn}{\nonumber \\}
\newcommand{\bra}{\langle}
\newcommand{\ket}{\rangle}
\newcommand{\fr}{\frac}

\begin{document}

\title{Orbital Dzyaloshinskii-Moriya Exchange Interaction}

\author{Panjin Kim}
\affiliation{Department of Physics and BK21 Physics Research
Division, Sungkyunkwan University, Suwon 440-746, Korea}
\author{Jung Hoon Han}
\email[Electronic address:$~~$]{hanjh@skku.edu}
\affiliation{Department of Physics and BK21 Physics Research
Division, Sungkyunkwan University, Suwon 440-746, Korea}
\affiliation{Asia Pacific Center for Theoretical Physics, POSTECH,
Pohang, Gyeongbuk 790-784, Korea}
\date{\today}

\begin{abstract} Superexchange calculation is performed for
multi-orbital band models with broken inversion symmetry.
Orbital-changing hopping terms allowed by the symmetry breaking
electric field lead to a new kind of orbital exchange interaction
closely resembling the Dzyaloshinskii-Moriya spin exchange.
Inversion symmetry breaking as present in surfaces and interfaces
and a strong on-site repulsion, but not the spin-orbit interaction,
are the requirements to observe the proposed effect. Mean-field
phase diagram exhibits a rich structure including
anti-ferro-orbital, ferro-orbital, and both single and multiple
spiral-orbital phases in close analogy with the Skyrmion spin
crystal phase recently discovered in thin-film chiral magnets.
\end{abstract}
\pacs{75.25.Dk, 75.30.Et, 75.10.Hk} \maketitle

\section{Introduction}

Strong on-site repulsion transforms the Bloch bands of nearly-free
electrons into an insulator where the residual low-energy dynamics
is that of the spin degrees of freedom interacting with each other
via the superexchange mechanism~\cite{anderson}. For
spin-orbit-coupled bands, the spin-flip hopping processes result in
another type of spin exchange called the Dzyaloshinskii-Moriya (DM)
interaction~\cite{DM} under the superexchange process.
Symmetry-wise, local inversion symmetry breaking such as the bond
distortion for a pair of adjacent magnetic orbitals, in addition to
the spin-orbit interaction (SOI), is the pre-requisite for the DM
interaction to make its appearance in a given system. Ordered
magnetic ground state is modified from being collinear as a result
of the DM exchange to favor spiral structure.

Meanwhile, inversion symmetry breaking (ISB) on the global scale
takes place for surfaces and interfaces and affects the band
structure with new effects such as the Rashba
interaction~\cite{rashba}. The situation was recently reviewed
carefully in Refs.~\onlinecite{hedegard,orbital-Rashba} where it was
shown that the symmetry-breaking electric field along the
surface-normal $z$-direction modifies the band structure within the
$xy$-plane by allowing previously forbidden hopping processes.
Examples are $p_{x(y)} \!\leftrightarrow\! p_z$ orbital hopping in
the $p$-band, and $d_{xy}\!\leftrightarrow\!d_{zx}$ orbital hopping
in the $t_{2g}$-band. It was further shown~\cite{orbital-Rashba}
that the new hopping terms arising from ISB can be cast in the form
$-\gamma \sum_{\v k} \Psi^\dag_{\v k} \v L \cdot (\v k \times
\hat{z} ) \Psi_{\v k}$ around the $\Gamma$ ($\v k=0$) point, where
$\Psi_{\v k}$ is the collection of, say, $p$-orbital operators
$(X_{\v k}, Y_{\v k}, Z_{\v k})^T$ in momentum coordinates $\v k$,
$\gamma$ is a parameter measuring the degree of ISB, and $\v L$ is
the spin-1 orbital angular momentum (OAM) operator. From the
structure of the new Hamiltonian it readily follows that each band
will carry polarized OAM proportional to $h (\v k \times \hat{z})$
with the respective helicities $h=+1, 0, -1$~\cite{orbital-Rashba}.
The enlargement of effective spin size from 1/2 (as in electrons'
spin) to 1 (as in degenerate $p$-orbital bands) results in the
appearance of the third band that remains unpolarized. The chiral
structure of the OAM, dubbed the ``orbital Rashba effect", can occur
even in the complete absence of SOI, and has been confirmed by
circular dichroism ARPES work on the weak-SOI material,
Cu~\cite{CD-ARPES}.

The new hopping processes allowed by ISB are in fact the orbital
analogues of spin-flip hoppings in spin-orbit-coupled bands.
Therefore, the two necessary conditions for the emergence of spin-DM
interaction - ISB and SOI - are both effectively fulfilled for
orbital magnetism when a symmetry-breaking electric field acts
perpendicular to the two-dimensional surface. The purpose of this
paper is to review this situation carefully in the limit of strong
on-site interaction regime to ask if an orbital analogue of spin-DM
interaction exists. In Sec. \ref{sec:SE-Hamiltonian} superexchange
calculation is carried out for multi-orbital tight-binding
Hamiltonian embodying the ISB. The emergence of the orbital DM
interaction is demonstrated in Sec. \ref{sec:phase-diagram} together
with the phase diagram exhibiting spiral and multi-spiral
structures. Possible observation of orbital DM-induced
orbital-spiral phases in magnetic thin films is discussed in Sec.
\ref{sec:summary}.

\section{Superexchange with ISB}
\label{sec:SE-Hamiltonian}

Assuming three degenerate $p$-orbital states at each site, a square
lattice Hamiltonian with nearest-neighbor hopping $H_t =
\sum_{i\sigma} ( H_{i,i+\hat{x},\sigma} + H_{i, i+\hat{y},\sigma} )$
is constructed,

\ba H_{i,i+\hat{x},\sigma} &=& t_a X_{i\sigma}^{\dag}
X_{i+\hat{x},\sigma} + t_b [Y_{i\sigma}^{\dag} Y_{i+\hat{x},\sigma}
\! + \!  Z_{i\sigma}^{\dag} Z_{i+\hat{x},\sigma} ] \nn
&& +\gamma ( X_{i\sigma}^{\dag} Z_{i+\hat{x},\sigma} \! -\!
Z_{i\sigma}^{\dag} X_{i+\hat{x},\sigma} ) + h.c. \nn
H_{i,i+\hat{y},\sigma} &=& t_a Y_{i\sigma}^{\dag}
Y_{i+\hat{y},\sigma} + t_b [X_{i\sigma}^{\dag} X_{i+\hat{y},\sigma}
\! +\!  Z_{i\sigma}^{\dag} Z_{i+\hat{y},\sigma} ] \nn
&& +\gamma ( Y_{i\sigma}^{\dag} Z_{i+\hat{y},\sigma} \! -\!
Z_{i\sigma}^{\dag} Y_{i+\hat{y},\sigma} ) + h.c. ~~.
\label{eq:Hij}\ea
Two hopping integrals $t_a$ and $t_b$ are introduced for $\sigma$-
and $\pi$-bonding orbital hoppings, respectively. Inter-orbital
hopping becomes possible when the ISB parameter $\gamma$ is
nonzero~\cite{orbital-Rashba,hedegard}. All hopping parameters are
real due to the assumed time-reversal invariance. Three-component
spinor can be formed, $\psi_i=\bpm X_i & Y_i & Z_i \epm^T$,
representing the $p_x$-, $p_y$-, and $p_z$-orbitals at the site $i$.
Multi-orbital Hubbard interaction is~\cite{oles}

\ba
H_U \!\!&=&\!\! U \!\sum\limits_{i\alpha}\! n_{i \alpha \uparrow} n_{i \alpha \downarrow}
    + \left(U-\fr{5}{2}J_H \right) \!\!\!\!\!\sum\limits_{i,\alpha < \beta, \sigma, \sigma'}\!\!\!\! n_{i \alpha \sigma} n_{i \beta \sigma'} \nn
    \!\!&-&\!\! 2 J_H \!\!\sum\limits_{i,\alpha < \beta}\! \mathbf{S}_{i \alpha} \! \cdot \! \mathbf{S}_{i \beta}
    +\! J_H \!\!\sum\limits_{i,\alpha \ne \beta}\! c_{i \alpha \uparrow}^{\dag} c_{i \alpha \downarrow}^{\dag} c_{i \beta \downarrow} c_{i \beta \uparrow},
\ea
where $U$ and $J_H$ are Coulomb and Hund's exchange elements,
$\alpha$, $\beta$ and $\sigma$, $\sigma'$ are orbital and spin
indices, respectively, and $n_{i\alpha\sigma}$ is a number operator
counting the electron number in $\alpha$-orbital with spin $\sigma$
at site $i$.

Shekhtman \textit{et al.} showed how to carry out the superexchange
calculation efficiently for spin-orbit-coupled bands by introducing
unitary rotations for operators to absorb spin-flip
hoppings~\cite{aharony}. We may adopt similar unitary rotations, in
the orbital subspace, to remove orbital-changing hoppings from the
Hamiltonian (\ref{eq:Hij}) with two unitary matrices

\ba
U_x \! = \! \left(\!
  \begin{array}{ccc}
    \cos\fr{\theta}{2} & 0 & \sin\fr{\theta}{2} \\
    0 & 1 & 0 \\
    -\sin\fr{\theta}{2} & 0 & \cos\fr{\theta}{2} \\
  \end{array}\!
\right)\!, ~U_y \! = \! \left(\!
  \begin{array}{ccc}
    1 & 0 & 0 \\
    0 & \cos\fr{\theta}{2} & \sin\fr{\theta}{2} \\
    0 & -\sin\fr{\theta}{2} & \cos\fr{\theta}{2} \\
  \end{array}\!
\right)\! . \ea
Two distinct rotations are required since the orbital-changing
hopping mixes $(p_x, p_z )$ orbitals along the $x$-direction, but
$(p_y, p_z)$ orbitals for the $y$-direction. As the new operators

\ba \tilde{\psi}_{i,x(y)} &=& U^\dag_{x(y)} \psi_i,\nn
\tilde{\psi}_{i+\hat{x}} &=& U_{x} \psi_{i+\hat{x}}, \nn
\tilde{\psi}_{i+\hat{y}} &=& U_{y} \psi_{i+\hat{y}}\ea
are inserted in Eq. (\ref{eq:Hij}) one obtains a new hopping
Hamiltonian $\tilde{H}_t = \sum_{i\sigma} (
\tilde{H}_{i,i+\hat{x},\sigma} + \tilde{H}_{i, i+\hat{y},\sigma})$,

\ba
\tilde{H}_{i,i+\hat{x},\sigma} &=& (t+t_2) \tilde{X}_{i\sigma}^{\dag}
\tilde{X}_{i+\hat{x},\sigma}+t_b \tilde{Y}_{i\sigma}^{\dag} \tilde{Y}_{i+\hat{x},\sigma} \nn
&&+(t-t_2) \tilde{Z}_{i\sigma}^{\dag} \tilde{Z}_{i+\hat{x},\sigma} + h.c. \nn
\tilde{H}_{i,i+\hat{y},\sigma} &=& t_b \tilde{X}_{i\sigma}^{\dag}
\tilde{X}_{i+\hat{x},\sigma} + (t+t_2) \tilde{Y}_{i\sigma}^{\dag}
\tilde{Y}_{i+\hat{x},\sigma} \nn
&&+(t-t_2) \tilde{Z}_{i\sigma}^{\dag} \tilde{Z}_{i+\hat{x},\sigma} + h.c.,
\ea
where $t(\cos \theta, \sin \theta) \!=\! (t_1, \gamma)$,
$t_1\!=\!(t_a \!+\! t_b)/2$, $t_2\!=\!(t_a \!-\! t_b)/2$. Although
bearing the same notation, the meaning of the tilde operators
appearing in $ \tilde{H}_{i,i+\hat{x},\sigma}$ is distinct from that
in $\tilde{H}_{i,i+\hat{y},\sigma}$ due to different sets of
rotations involved.

Superexchange calculation at 1/6-filling (one particle per site) can
proceed now via standard methods with the Hamiltonian $H=
\tilde{H}_t + H_U$, where the orbital-changing hopping terms are
seemingly absent. The exchange Hamiltonian thus obtained, writing
$t+t_2 \equiv t_{\tilde{X}}$, $t_b \equiv t_{\tilde{Y}}$ and $t-t_2
\equiv t_{\tilde{Z}}$, reads ${\cal H}={\cal H}_{\hat{x}} + {\cal
H}_{\hat{y}}$,

\ba {\cal H}_{\hat{x}} \!&=&\! J_{T_1} \sum_i \Big(\v S_i \cdot \v
S_{i+\hat{x}} \!+\! \fr{3}{4}\Big) \Big(\hat{A}_{i,i+\hat{x}} + \hat{B}_{i,i+\hat{x}} \!-\! \hat{N}_{i,i+\hat{x}}\Big)\! \nn
&+&\! J_{E} \sum_i \Big(\v S_i \cdot \v
S_{i+\hat{x}} \!-\! \fr{1}{4}\Big) \Big(\!\!-\!\hat{A}_{i,i+\hat{x}} + \hat{B}_{i,i+\hat{x}} \!+\! \hat{N}_{i,i+\hat{x}}\Big)\! \nn
&+&\! J_{A_1} \sum_i \Big(\v S_i \cdot \v
S_{i+\hat{x}} \!-\! \fr{1}{4}\Big) \Big(\fr{2}{3}\hat{A}_{i,i+\hat{x}} \!+\! \fr{2}{3}\hat{C}_{i,i+\hat{x}}\Big)\! \nn
&+&\! J_{T_2} \sum_i \Big(\v S_i \cdot \v
S_{i+\hat{x}} \!-\! \fr{1}{4}\Big) \Big(\fr{4}{3}\hat{A}_{i,i+\hat{x}} \!-\! \fr{2}{3}\hat{C}_{i,i+\hat{x}}\Big),\!
 \label{eq:spin-orbital-exchange}\ea
where $J_{T_1} = 2/(U-3J_H)$, $J_{E} = J_{T_2} = 2/(U-J_H)$, and
$J_{A_1} = 2/(U+2J_H)$. Orbital exchange parts are given by

\ba &&\hat{A}_{i,i+\hat{x}} = \sum_{\alpha} t_{\alpha}^2 n_{i\alpha} n_{j\alpha}    , \nn
&&\hat{B}_{i,i+\hat{x}} = \sum_{\alpha < \beta} t_{\alpha} t_{\beta} (\alpha^\dag_i \beta_i \beta_{i+\hat{x}}^\dag \alpha_{i+\hat{x}}
+  \beta^\dag_i \alpha_i \alpha_{i+\hat{x}}^\dag \beta_{i+\hat{x}} )   , \nn
&&\hat{C}_{i,i+\hat{x}} = \sum_{\alpha < \beta} t_{\alpha} t_{\beta} (\alpha^\dag_i \beta_i \alpha_{i+\hat{x}}^\dag \beta_{i+\hat{x}}
+  \beta^\dag_i \alpha_i \beta_{i+\hat{x}}^\dag \alpha_{i+\hat{x}} ), \nn
&&\hat{N}_{i,i+\hat{x}} = \fr{1}{2}\sum_{\alpha, \beta} (t_{\alpha}^2 n_{i\alpha} + t_{\beta}^2 n_{i+\hat{x},\beta})   ,
 \label{eq:orbital-exchange} \ea
where $\alpha_i$ $(\alpha_i^\dag)$ annihilates (creates) electrons
in the $\alpha$-orbital at $i$-site. ${\cal H}_{\hat{y}}$ is easily
obtained from ${\cal H}_{\hat{x}}$ by switching $t_{\tilde{X}}
\leftrightarrow t_{\tilde{Y}}$ and replacing $i+\hat{x}$ by
$i+\hat{y}$. This Hamiltonian resembles the spin-orbital model
describing $\text{LaTiO}_3$ system studied by Khaliullin and his
colleagues~\cite{khaliullin} in the sense that in $\text{LaTiO}_3$
system, each site has one active electron occupying one of three
$t_{2g}$-orbitals. Yet, there are differences coming from the fact
that while our model (\ref{eq:spin-orbital-exchange}) after the
rotation is dealing with three unequal hopping integrals between
adjacent orbitals, in $\text{LaTiO}_3$ system the $\pi$-hopping is
ignored and the other two integrals have equal
strengths~\cite{khaliullin}. By ignoring $\pi$-hopping and taking
the limit $t+t_2 = t-t_2$ in Eq. (\ref{eq:spin-orbital-exchange}),
our superexchange Hamiltonian becomes identical to that of
$\text{LaTiO}_3$. Equations (\ref{eq:spin-orbital-exchange}) and
(\ref{eq:orbital-exchange}) constitute the main technical findings
of the present work.

Some comments about the limiting cases are in order. The exchange
Hamiltonian (\ref{eq:spin-orbital-exchange}) simplifies greatly in
the $J_H = 0$ limit,

\ba {\cal H}_{\hat{x}} \!&=&\! 4 \sum_i \Big(\v S_i \cdot \v
S_{i+\hat{x}} \!+\! \fr{1}{4}\Big) \sum_{\alpha,\beta}  t_\alpha
t_\beta \alpha_i^{\dag} \beta_i \beta_{i+\hat{x}}^{\dag}
\alpha_{i+\hat{x}} \! \nn
&-& \!\sum_{i, \alpha} t_\alpha^2 (n_{i \alpha} \!+\! n_{i+\hat{x},
\alpha}), \nn
{\cal H}_{\hat{y}} \!&=&\! {\cal H}_{\hat{x}}|{(t_{\tilde{X}}
\leftrightarrow t_{\tilde{Y}},\; i+\hat{x} \rightarrow
i+\hat{y})}, \label{eq:spin-orbital-exchange2}\ea
where $U$ is taken to be unity. Furthermore, in the isotropic
limit $t_{\tilde{X}} = t_{\tilde{Y}} = t_{\tilde{Z}}$ the pairwise
exchange interaction in Eq. (\ref{eq:spin-orbital-exchange2})
possesses the SU(6) symmetry in the combined spin and orbital
spaces~\cite{SU(4)}. The larger symmetry can be most easily seen by
defining operators $\Psi_{i,\sigma \alpha} = \varphi_{i\sigma}
\alpha_i$ and $\Psi_{i,\sigma \alpha}^\dag = \alpha_i^\dag
\varphi_{i\sigma}^\dag $, where $\varphi_{i\sigma}$
$(\varphi_{i\sigma}^{\dag})$ annihilates (creates) electrons with
spin $\sigma$ at $i$-site. The pairwise exchange Hamiltonian can be
re-written in the manifestly SU(6)-invariant form ${\cal H}_{ij}
\sim \sum_{K,K'} \Psi_{j,K}^\dag \Psi_{i,K} \Psi_{i,K'}^\dag
\Psi_{j,K'}$, where $K$ and $K'$ run over six possible spin and
orbital configurations. The SU(6) symmetry will remain for
one-dimensional chain consisting of either ${\cal H}_{\hat{x}}$ or
${\cal H}_{\hat{y}}$ alone, but not for the two-dimensional model
${\cal H}_{\hat{x}} + {\cal H}_{\hat{y}}$ due to the fact that two
different sets of unitary rotations were used to arrive at the
overall superexchange Hamiltonian, Eq.
(\ref{eq:spin-orbital-exchange2}).

\section{Orbital DM and Phase diagram}
\label{sec:phase-diagram}
In this section we explicitly point out the emergence of DM-type
orbital exchange interaction in our model and study possible phase
diagram using the site-factorization scheme. DM-type interactions
will be recovered by un-rotating the Hamiltonian
(\ref{eq:orbital-exchange}) to the original orbital basis as shown
in Ref. \onlinecite{aharony}. In doing so for our Hamiltonian
(\ref{eq:spin-orbital-exchange}) one encounters unwieldy expressions
that simplify somewhat by taking the weaker $\pi$-bonding to zero:
$t_b =0$. This limit is usually taken in the superexchange
calculation for $d$- orbital systems in transition metal
oxides~\cite{khaliullin}, and more recently for $p$-orbital systems
in an optical lattice~\cite{lewenstein}.

In such limit the orbital exchange operators in Eq.
(\ref{eq:orbital-exchange}) can be re-expressed in terms of
Gell-Mann matrices $\lambda^\alpha$ ($\alpha=1\cdots8$) as follows:

\begin{widetext}
\ba
{\cal H}_{\hat{x}} \!&=&\! \fr{J_{T_1}}{4}\sum_i \Big(\v S_i \cdot \v S_{i+\hat{x}} + \fr{3}{4} \Big) \Big(t_{\tilde{X}}^2 [\lambda_i^x + (\lambda_i^x)^2][\lambda_{i+\hat{x}}^x + (\lambda_{i+\hat{x}}^x)^2] + t_{\tilde{Z}}^2 [-\lambda_i^x + (\lambda_i^x)^2][-\lambda_{i+x}^x + (\lambda_{i+x}^x)^2]  \nn
   && ~~~~~~~~~~~ + 2 t_{\tilde{X}} t_{\tilde{Z}} [\lambda_i^4 \lambda_{i+\hat{x}}^4 + \lambda_i^5 \lambda_{i+\hat{x}}^5 ] \Big)   \nn
   &-&\fr{J_{E}}{4}\sum_i \Big(\v S_i \cdot \v S_{i+\hat{x}} - \fr{1}{4} \Big) \Big(t_{\tilde{X}}^2 [\lambda_i^x + (\lambda_i^x)^2][\lambda_{i+\hat{x}}^x + (\lambda_{i+\hat{x}}^x)^2] + t_{\tilde{Z}}^2 [-\lambda_i^x + (\lambda_i^x)^2][-\lambda_{i+x}^x + (\lambda_{i+x}^x)^2]      \nn
   && ~~~~~~~~~~~ - 2 t_{\tilde{X}} t_{\tilde{Z}} [\lambda_i^4 \lambda_{i+\hat{x}}^4 + \lambda_i^5 \lambda_{i+\hat{x}}^5 ] \Big)   \nn
   &+&\fr{J_{A_1}}{3}\sum_i \Big(\v S_i \cdot \v S_{i+\hat{x}} - \fr{1}{4} \Big) \Big(\fr{1}{2} t_{\tilde{X}}^2 [\lambda_i^x + (\lambda_i^x)^2][\lambda_{i+\hat{x}}^x + (\lambda_{i+\hat{x}}^x)^2] + \fr{1}{2} t_{\tilde{Z}}^2 [-\lambda_i^x + (\lambda_i^x)^2][-\lambda_{i+x}^x + (\lambda_{i+x}^x)^2]  \nn
   && ~~~~~~~~~~~ + t_{\tilde{X}} t_{\tilde{Z}} [\lambda_i^4 \lambda_{i+\hat{x}}^4 - \lambda_i^5 \lambda_{i+\hat{x}}^5] \Big)   \nn
   &+&\fr{J_{T_2}}{3}\sum_i \Big(\v S_i \cdot \v S_{i+\hat{x}} - \fr{1}{4} \Big) \Big(t_{\tilde{X}}^2 [\lambda_i^x + (\lambda_i^x)^2][\lambda_{i+\hat{x}}^x + (\lambda_{i+\hat{x}}^x)^2] + t_{\tilde{Z}}^2 [-\lambda_i^x + (\lambda_i^x)^2][-\lambda_{i+x}^x + (\lambda_{i+x}^x)^2]  \nn
   && ~~~~~~~~~~~ - t_{\tilde{X}} t_{\tilde{Z}} [\lambda_i^4 \lambda_{i+\hat{x}}^4 - \lambda_i^5 \lambda_{i+\hat{x}}^5] \Big)   \nn
   &-& t_{\tilde{X}}^2(n_{i, \tilde{X}} + n_{i+\hat{x}, \tilde{X}}) - t_{\tilde{Z}}^2(n_{i, \tilde{Z}} + n_{i+\hat{x}, \tilde{Z}}), \nn \nn
{\cal H}_{\hat{y}} \!&=&\! {\cal H}_{\hat{x}} | (t_{\tilde{X}} \!\rightarrow\! t_{\tilde{Y}},\; \lambda^x
\!\rightarrow\! \lambda^y,\; \lambda^4 \!\rightarrow\!
\lambda^6,\; \lambda^5 \!\rightarrow\! \lambda^7,\; i\!+\!\hat{x}
\rightarrow i\!+\!\hat{y},\; n_{i, \tilde{X}} \!\rightarrow\! n_{i, \tilde{Y}},\; n_{i+\hat{x}, \tilde{X}} \!\rightarrow\! n_{i+\hat{y}, \tilde{Y}}). \label{eq:Hx-Hy} \ea
\end{widetext}
For convenience we introduced $\lambda^x = \mathrm{diag} (1, 0, -1)$
and $\lambda^y = \mathrm{diag} (0, 1, -1)$ as combinations of
$\lambda^3$, $\lambda^8$ and the unit matrix. It is useful to note
that ${\cal H}_{\hat{x}}$ is entirely constructed from three
matrices, $\v T^x_i =(\lambda^4_i , \lambda^5_i , \lambda^x_i )$,
reducing to a set of Pauli matrices in the orbital subspace of $(p_x
, p_z)$. Similarly, ${\cal H}_{\hat{y}}$ employs another set of
three matrices and $\v T^y_i = (\lambda^6_i , \lambda^7_i ,
\lambda^y_i )$ reducing to Pauli matrices in the $(p_y, p_z)$
orbital subspace.

Returning to the original basis $(p_x , p_y, p_z )$ amounts to
making the unitary replacements

\ba (\lambda_i^\alpha , \lambda_{i+\hat{x}}^\alpha ) &\rightarrow&
(U_x \lambda_i^\alpha U^\dag_x , U^\dag_x \lambda_{i+\hat{x}}^\alpha
U_x ), \nn
(\lambda_i^\alpha , \lambda_{i+\hat{y}}^\alpha ) & \rightarrow &
(U_y \lambda_i^\alpha U^\dag_y , U^\dag_y \lambda_{i+\hat{y}}^\alpha
U_y ), \ea
in Eq. (\ref{eq:Hx-Hy}). After the rotation there appear terms
linear in $\gamma$,
\ba t \gamma \sum_i \Bigl( [\lambda_i^x + (\lambda_i^x)^2
]\lambda_{i+\hat{x}}^4 - \lambda_i^4 [\lambda_{i+\hat{x}}^x + (
\lambda_{i+\hat{x}}^x )^2 ] \nn
+ [\lambda_i^y + (\lambda_i^y)^2
]\lambda_{i+\hat{y}}^6 - \lambda_i^6 [ \lambda_{i+\hat{y}}^y +
(\lambda_{i+\hat{y}}^y)^2 ] \Bigr) . \label{eq:3orbital-DM} \ea
As mentioned earlier, $(\lambda^4, \lambda^5, \lambda^x )$ are
effectively replaced by the Pauli matrices $\bm \tau = (\tau^x,
\tau^y, \tau^z)$ within the $(p_x, p_z )$-orbital subspace, thus the
first line of Eq. (\ref{eq:3orbital-DM}) becomes

\ba t \gamma \sum_i [ \tau_i^z \tau_{i+\hat{x}}^x -\tau_i^x
\tau_{i+\hat{x}}^z ] = t \gamma\sum_i \hat{y}\cdot [ \bm \tau_i
\times \bm \tau_{i+\hat{x}} ] .\label{eq:orbital-DM}\ea
This is the orbital analogue of the DM spin exchange, or ``orbital
DM" (ODM) exchange. The real-valued transition amplitudes obtained
from the superexchange process necessarily excludes the imaginary
$\tau_y$ operator, permitting $\hat{y}\cdot [ \bm \tau_i \times \bm
\tau_{i+\hat{x}} ]$ as the only permissible form of DM interaction.
Analogously, the second line of Eq. (\ref{eq:3orbital-DM}) reduces
to the two-component $(p_y, p_z)$-orbital model with $(\lambda^6,
\lambda^7, \lambda^y)$ acting as another set of Pauli matrices $\bm
\mu$ with the DM interaction, $2 t \gamma \sum_i \hat{y} \cdot [\bm
\mu_i \times \bm \mu_{i+\hat{y}}]$. We emphasize that such Pauli
matrix description (effective spin-1/2 model) must give way to the
full Gell-Mann matrix formalism shown in Eq. (\ref{eq:Hx-Hy}) once
${\cal H}_{\hat{x}}$ and ${\cal H}_{\hat{y}}$ are combined in the
two-dimensional lattice.

A similar Gell-Mann matrix expression appeared in the low-energy
theory of hard-core three-component bosons confined in the optical
lattice~\cite{lewenstein} where, however, the influence of ISB on
the superexchange process was not examined. Superexchange
calculation for the $t_{2g}$-orbitals in the presence of
GdFeO$_3$-type distortion was carried out by Ishihara \emph{et
al.}~\cite{ishihara}. The oxygen distortions assumed in their work is
staggered, in the sense that the net displacement vector of all the
oxygen atoms is zero. On the contrary, we are dealing with the
situation where displacement of the $p_z$ is uniform, due to
external fields. Orbital analogue of the DM exchange as shown here
might be anticipated on symmetry grounds, but has never been
explicitly demonstrated before.

In proceeding to the mean-field analysis of the possible phases of
our Hamiltonian we assume that $J_\mathrm{H}$ is sufficiently large
to favor the ferromagnetic spin state~\cite{khaliullin}, $\bra \v
S_i \cdot \v S_j \ket = 1/4$. The approximation allows us to focus
on the orbital sector, which makes Eq. (\ref{eq:Hx-Hy}) simple
enough to be written down in original basis $\psi_i=\bpm X_i & Y_i &
Z_i \epm^T$ as

\begin{widetext}
\ba && {\cal H}_{\hat{x}}=2 t^2 \sum_i [\lambda_i^x +(\lambda_i^x
)^2 -1][ \lambda_{i+\hat{x}}^x + (\lambda_{i+\hat{x}}^x)^2 -1] + 2t
\gamma \sum_i \Bigl( [\lambda_i^x + (\lambda_i^x)^2
]\lambda_{i+\hat{x}}^4
      -\lambda_i^4 [\lambda_{i+\hat{x}}^x + ( \lambda_{i+\hat{x}}^x )^2 ]\Bigr) \nn
&&~~~~~~~    + \gamma^2 \sum_i \Bigl( \lambda_i^5
\lambda_{i+\hat{x}}^5 -
      \lambda_i^4 \lambda_{i+\hat{x}}^4 - \lambda_i^x \lambda_{i+\hat{x}}^x +
      [ (\lambda_i^x )^2 -1 ][ (\lambda_{i+\hat{x}}^x)^2 -1 ] \Bigr) , \nn
      &&  {\cal H}_{\hat{y}}=2 t^2 \sum_i [ \lambda_i^y +(\lambda_i^y )^2  -1] [
\lambda_{i+\hat{y}}^y+(\lambda_{i+\hat{y}}^y )^2 -1 ]+ 2 t \gamma
\sum_i \Big([\lambda_i^y + (\lambda_i^y)^2 ]\lambda_{i+\hat{y}}^6
-\lambda_i^6 [ \lambda_{i+\hat{y}}^y + (\lambda_{i+\hat{y}}^y)^2 ]
\Bigr) \nn
&&~~~~~~~    + \gamma^2 \sum_i \Bigl( \lambda_i^7
\lambda_{i+\hat{y}}^7 -
      \lambda_i^6 \lambda_{i+\hat{y}}^6 - \lambda_i^y \lambda_{i+\hat{y}}^y+
      [ (\lambda_i^y )^2 -1 ][ (\lambda_{i+\hat{y}}^y)^2 -1 ] \Bigr), \label{eq:Hx-Hy2}\ea
\end{widetext}
where $U-3J_H$ is taken to be unity and $t \equiv t_a/2$ for
brevity. The Hamiltonian (\ref{eq:Hx-Hy2}) is replaced by its
mean-field form by making the on-site ansatz of the wave function,

\ba |\psi_i \rangle = A^x_i |X_i \rangle + A^y_i |Y_i \rangle +
A^z_i |Z_i \rangle , \ea
with three complex coefficients satisfying $|A^x_i |^2 + |A^y_i |^2
+ |A^z_i |^2 =1$. The many-body wave function becomes the direct
product $|\psi\rangle = \prod_i \otimes |\psi_i \rangle$, and one
can make replacement of the operator $\lambda_i^\alpha$ with its
average, $\langle \lambda_i^\alpha \rangle$, in the Hamiltonian. In
this scheme one can choose $A^x_i$ real without loss of generality
and parameterize the coefficients generally as $A^x_i = \cos
\alpha_i $, $A^y_i = e^{i\gamma_i } \sin \alpha_i \cos \beta_i $,
$A^z_i = e^{i\delta_i } \sin \alpha_i \sin \beta_i $. The mean-field
Hamiltonian being a function of the four angles $(\alpha_i, \beta_i,
\gamma_i, \delta_i )$ per site can be minimized by the Monte Carlo
(MC) annealing method. We may enrich the phase diagram of the model
somewhat by introducing one more parameter $A$ to replace $t \gamma
\rightarrow t A \gamma$. Various renormalization of the interaction
parameters may alter the coefficients in the superexchange
Hamiltonian from those shown in Eq.
(\ref{eq:spin-orbital-exchange}), which we attempt to model with the
extra parameter $A$. It is also an attempt to understand the
possible phases of the orbital DM model within the wider perspective
than might be allowed from second-order perturbation.

\begin{figure}[htb]
    \centering
    \includegraphics[width=0.45\textwidth]{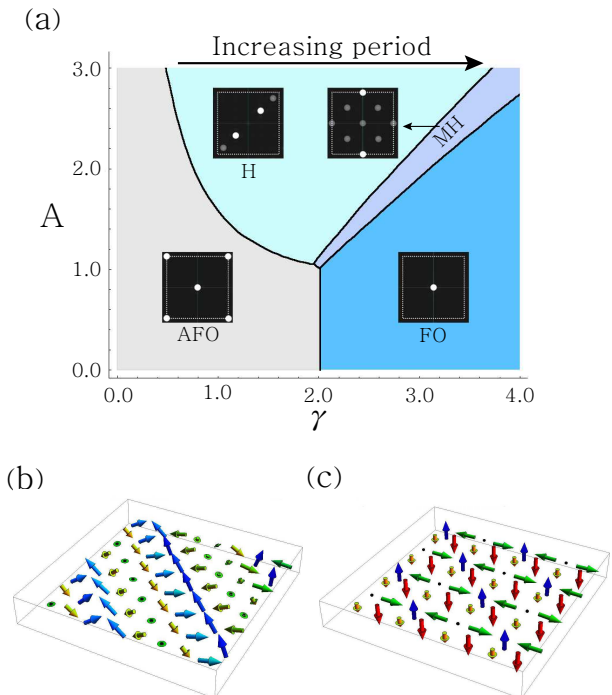}
    \caption{(color online) (a) Zero-temperature
    phase diagram of the recovered orbital model in Eq. (\ref{eq:Hx-Hy2}) with
    $t=1$.
    Phase boundaries are obtained on the basis of MC simulation and variational
    energy calculation. Orbital configurations are abbreviated as AFO
    (antiferro-orbital), FO (ferro-orbital), H (helical), and MH (multiple helical).
    Bragg patterns for each phase are schematically shown with dotted line denoting
    the first Brillouin zone boundary. In H, the period increases continuously
    as $\gamma$ becomes larger. Bragg spots at $(k,k)$ and $(3k,3k)$ are
    both present in H. Real space spin configurations in terms of
    $\langle \v T^x_i \rangle$ for (b) H and (c) MH phases are depicted.}
    \label{fig:phasediagram}
\end{figure}

The zero-temperature phase diagram spanned by $(\gamma, A)$ with $t$
fixed to unity is shown in Fig. \ref{fig:phasediagram}. Six
Gell-Mann matrices, organized into two groups $\v T^x_i
=(\lambda^4_i , \lambda^5_i , \lambda^x_i )$ and $\v T^y_i =
(\lambda^6_i , \lambda^7_i , \lambda^y_i )$, are used to
characterize the ground states by following their averages and their
Fourier components: $\v T_{\v k}^n = \sum_{i} \bra \v T_{i}^n \ket
e^{- i \v k \cdot \v r_i }$ ($n=x,y$). The Fourier analysis is
particularly helpful in searching for modulated structures with long
periods that often appear with the DM interaction.

When $\gamma=0$ the Hamiltonian is expressed entirely in terms of
two commuting matrices, ($\lambda^x_i, \lambda^y_i )$, with the
ground state given by alternate occupations of $p_x$ and $p_y$
orbitals on the square lattice.  This phase, called the
antiferro-orbital (AFO) state, dominates the small-$\gamma$ region
of the phase diagram. For $A$ small and $\gamma$ increasing beyond a
critical value, one finds a first-order transition into a
ferro-orbital (FO) state with $p_z$-orbital occupation at every
site. In the phase diagram of Fig. \ref{fig:phasediagram} one finds
that we are considering rather large $\gamma$ values. While this may
be unlikely in conventional solid materials, a thin film consisting
of narrow-band materials subject to very large perpendicular
electric field may be able to realize such physical regime.
Inhabiting the large-$A$, intermediate-$\gamma$ region is the
$\langle 11 \rangle$ helical (H) phase that we found to be well
described by

\ba \langle \v T_i^x \rangle \!\!\!&=&\!\!\! \Big(\!\! \sin 2[ \v k
\cdot \v r_i\!+\!\alpha_0 ] \sin[ \v k \cdot \v r_i\!+\!\beta_0 ] ,
                    0,      \nn
&&       \!\!\!    \cos^2[ \v k \cdot \v r_i+ \alpha_0 ] - \sin^2[
\v k \cdot \v r_i\!+\!\alpha_0]  \sin^2 [ \v k \cdot \v r_i \!+\!
\beta_0 ] \Big), \nn
\langle \v T_i^y \rangle \!\!\!&=&\!\!\! \Big(\!\! \sin^2[ \v k
\cdot \v r_i\!+\!\alpha_0 ]  \sin 2[ \v k \cdot \v r_i\!+\! \beta_0
], 0 , \nn &&      ~~~~  \sin^2[ \v k \cdot \v r_i\!+\!\alpha_0 ]
\cos 2[ \v k \cdot \v r_i\!+\!\beta_0 ] \Big), \ea
$ \v k = (k,k)$. Variational calculation of the minimum energy with
respect to $k$ and the relative phase angle $\alpha_0 -\beta_0$
confirms that the period $2\pi/k$ is increased with $\gamma$, 2 on
the smaller $\gamma$ side to $>6$, before it is supplanted by another
intricate phase taking place between H and FO. The emergence of
orbital spiral phase is a natural consequence of the orbital DM
exchange.

The new phase, indicated as MH in Fig. \ref{fig:phasediagram}, is
constructed as the equal-weight superposition of two helices with
$\v k_1 = \pm(\pi/2,\pi/2)$ and $\v k_2 = \pm (\pi/2,-\pi/2)$, as
well as two pairs of peaks at $\pm (\pi,0)$ and $\pm (0,\pi)$. The
intensities obtained from Fourier analysis of $\langle \v T_i^x
\rangle $ (inset in Fig. \ref{fig:phasediagram}) show stronger peaks
at $\pm(0,\pi)$ than at $\pm(\pi,0)$. On the other hand Fourier
analysis of $\langle \v T_i^y \rangle$ revealed the Bragg peaks at
$\pm (\pi,0)$ are brighter than at $\pm (0,\pi)$. One can still draw
a close parallel of the MH phase found in the present model to the
square lattice of Skyrmions and anti-Skyrmions found in some models
of spiral magnetism~\cite{bogdanov,yi}, which also consists of
multiple Bragg peaks at $\pm( k, k)$ and $\pm (k,-k)$ in its spin
structure.

\section{Conclusion and Summary}
\label{sec:summary}

Orbital ordering in multi-band Hubbard systems have been studied for
several decades since the pioneering work of Kugel and Khomskii
(KK)~\cite{KK}. Extension of the original two-orbital KK model to
three-orbital $t_{2g}$ case has been thoroughly carried out by
Khaliullin and collaborators~\cite{khaliullin}. Recent works on the
optical lattice of cold atoms also arrived at three-orbital exchange
model~\cite{lewenstein}, without the spin degrees of freedom.
Meanwhile, remarkable advances in the growth technique of ultrathin
materials prompt consideration of the influence of ISB, $\gamma \neq
0$, on the electronic band structure and, as we discuss in this
paper, on the orbital physics as well. With this background, we have
derived the analogue of spin-DM exchange interaction as a natural
consequence of ISB in the multi-orbital Hubbard model. Physical
requirements for its appearance are the multi-orbital degeneracy and
the loss of inversion symmetry, but not the spin-orbit interaction
as in the spin-DM exchange.

Although the derivations presented in this paper are based on the
$p$-orbital picture, the case of degenerate $t_{2g}$-orbitals with
ISB can be worked out, without further calculation, by making the
replacements $(p_x, p_y, p_z ) \rightarrow (d_{yz}, d_{zx},
d_{xy})$, and switching $\sigma$- $\leftrightarrow$ $\pi$-hopping
integrals $t_a \leftrightarrow t_b$ in all our results. It is thus
expected that conclusions regarding the phase diagram as shown in
Fig. \ref{fig:phasediagram} may be directly applicable to ultra-thin
films made of transition-metal elements. Unlike the spin-DM
interaction in materials, the governing factor $\gamma$ responsible
for the orbital-DM exchange can be imposed externally by the electric
field in a controlled manner. Interesting quantum-orbital phases and
transitions between them may be observed in a thin-film
multi-orbital system subject to perpendicular electric field of
variable strength.

\acknowledgments J. H. H. is supported by NRF grant (No.
2011-0015631). P. K. is supported by NRF grant funded by the Korean
Government (NRF-2012) - Global Ph. D. Fellowship Program. Insightful
comments from Giniyat Khaliullin and Hosho Katsura are gratefully
acknowledged.


\begin{thebibliography}{14}

\bibitem{anderson} P. W. Anderson, Phys. Rev. \textbf{79}, 350 (1950).

\bibitem{DM} I. E. Dzyaloshinskii, J. Chem. Solids \textbf{4}, 241 (1958);
T. Moriya, Phys. Rev. \textbf{120}, 91 (1960); Phys. Rev. Lett.
\textbf{4}, 228 (1960).

\bibitem{rashba} Y. A. Bychkov and E. I. Rashba, JETP Lett. {\bf 39}, 78
(1984).

\bibitem{hedegard} L. Petersen and P. Hedeg{\aa}rd, Surf. Sci.
\textbf{459}, 49 (2000).

\bibitem{orbital-Rashba} Jin-Hong Park, Choong H. Kim, Jun-Won Rhim, and Jung Hoon Han,
Phys. Rev. B \textbf{85}, 195401 (2012).

\bibitem{CD-ARPES} Beomyoung Kim, Choong H. Kim, Panjin Kim, Wonsig Jung,
Yeongkwan Kim, Yoonyoung Koh, Masashi Arita, Kenya Shimada, Hirofumi
Namatame, Masaki Taniguchi, Jaejun Yu, and Changyoung Kim, Phys.
Rev. B \textbf{85}, 195402 (2012).

\bibitem{oles} A. M. Ole$\acute{\text{s}}$,
Phys. Rev. B \textbf{28}, 327 (1983).

\bibitem{aharony} L. Shekhtman, O. Entin-Wohlman,
and Amnon Aharony, Phys. Rev. Lett. \textbf{69}, 836 (1992).

\bibitem{khaliullin} G. Khaliullin and S. Maekawa, Phys. Rev. Lett. \textbf{85}, 3950
(2000); G. Khaliullin and S. Okamoto, Phys. Rev. B \textbf{68},
205109 (2003); G. Khaliullin, Prog. Theor. Phys. Suppl.
\textbf{160}, 155 (2005).

\bibitem{SU(4)} Daniel P. Arovas and Assa Auerbach, Phys. Rev. B \textbf{52}, 10114
(1995); Y. Q. Li, M. Ma, D. N. Shi, and F. C. Zhang, Phys. Rev.
Lett. \textbf{81}, 3527 (1998); Y. Yamashita, N. Shibata, and K.
Ueda, Phys. Rev. B \textbf{58}, 9114 (1998); Swapan K. Pati, Rajiv
R. P. Singh, and Daniel I. Khomskii, Phys. Rev. Lett. \textbf{81},
5406 (1998).

\bibitem{lewenstein} Philipp Hauke, Erhai Zhao, Krittika Goyal,
Ivan H. Deutsch, W. Vincent Liu, and Maciej Lewenstein, Phys. Rev. A
\textbf{84}, 051603(R) (2011).

%


\bibitem{ishihara} S. Ishihara, T. Hatakeyama, and S. Maekawa,
Phys. Rev. B \textbf{65}, 064442 (2002).

\bibitem{KK} K. I. Kugel and D. I. Khomskii, Sov. Phys. Usp. \textbf{25}, 231
(1982).

\bibitem{bogdanov} U. K. Ro{\ss}ler, A. N. Bogdanov, and C. Pfleiderer,
Nature \textbf{442}, 797 (2006).

\bibitem{yi} Su Do Yi, Shigeki Onoda, Naoto Nagaosa, and Jung Hoon
Han, Phys. Rev. B \textbf{80}, 054416 (2009).


\end{thebibliography}
\end{document}